\documentclass[aps, prd, superscriptaddress, preprintnumbers, showpacs]{revtex4}
\usepackage[english]{babel}
\usepackage{graphicx,amsfonts,amsmath,amssymb,amstext}
\usepackage{float,wrapfig}
\usepackage{subfigure, psfrag}
\usepackage{dsfont}
\usepackage{epsfig}
\usepackage{color}
\usepackage{bm}
\usepackage{multirow,textcomp}
\usepackage{natbib}
\usepackage[normalem]{ulem}
\usepackage{hyperref}

\newcommand{\be}{\begin{equation}}
\newcommand{\ee}{\end{equation}}
\newcommand{\ba}{\begin{eqnarray}}
\newcommand{\ea}{\end{eqnarray}}

\definecolor{purple}{rgb}{0.8,0,0.6}

\makeatletter
\newcommand{\vast}{\bBigg@{2}}
\newcommand{\Vast}{\bBigg@{3}}
\makeatother

\begin{document}
\title{The Schwinger effect by axial coupling in  natural inflation model}
\date{\today}

\author{Mehran  Kamarpour}
\affiliation{Physics Faculty, Taras Shevchenko National University of Kyiv, 64/13, Volodymyrska str., 01601 Kyiv, Ukraine. Email: mehrankamarpour@yahoo.com}

\begin{abstract}
	We investigate the process of the Schwinger effect by axial coupling  in the natural single-field inflation model in two parts. First we consider the Schwinger effect when the conformal invariance of Maxwell action should be  broken by axial coupling $ I(\phi)F_{\mu\nu}\tilde{F}^{\mu\nu} $ with the inflaton field by identifying the standard horizon scale $  k=aH  $ at the very beginning of inflation  for additional boundary term and use  several values of coupling constant $ \chi_{1} $ and estimate electric and magnetic energy densities and energy density of produced charged particles due to the Schwinger effect.We find that for both coupling functions the energy density of the produced charged particles due to the Schwinger effect is so high and spoils inflaton field.In fact the strong coupling or back-reaction occurs because the energy density of produced charged particles is exceeding of inflaton field.We use two coupling functions to break conformal invariance of maxwell action.The simplest coupling function $ I\left(\phi\right)=\chi_{1}\frac{\phi}{M_{p}} $  and a curvature based  coupling function $ I\left(\phi\right)= 12\chi_{1}e^{\left(\sqrt{\frac{2}{3}}\frac{\phi}{M_{p}}\right)}\left[\frac{1}{3M_{p}^{2}}\left(4V\left(\phi\right)\right)+\frac{\sqrt{2}}{\sqrt{3}M_{p}}\left(\frac{dV}{d\phi}\right)\right] $ where $V\left(\phi\right)  $ is the potential of natural inflation. In second part , in oder to avoid strong back-reaction problem we identify the horizon scale $ k_{H}=aH|\zeta| ,      \zeta=\frac{{I}^{\prime}\left(\phi\right)\dot{\phi}}{H} $ in which a given Fourier begins to become tachyonically unstable.The effect of this scale is reducing the value of coupling constant $ \chi_{1} $ and weakening the back-reaction problem but in both cases strong coupling  or strong back-reaction exists and the Schwinger effect is impossible. Therefore, the Schwinger effect in this model is not possible and spoils inflation.Instantly,the Schwinger effect produces very high energy density of charged particles which causes back-reaction problem and spoils inflaton field.We must stress that due to existence of strong back-reaction in two cases the energy density of the produced charged particles due to the Schwinger effect spoils inflaton field and we do not reach to the so-called conductivity of plasma.

\end{abstract}


\maketitle

\tableofcontents

\section{Introduction}
\label{sec-intro}

We have recently shown in the natural single-field  inflation model with kinetic coupling both magneto-genesis and the Schwinger effect exist.\cite{Kamarpour-Sobol:2018,Kamarpour:2023}.In Refs.\cite{Kronberg:1994,Grasso:2001,Widrow:2002,Giovannini:2004,Kandus:2011,Durrer:2013,Subramanian:2016} have been shown the strength of  detected magnetic    fields indicates a wide range,with values ranging from a few micogauss in galaxies and in cluster of galaxies to a very high as $ 10^{15} $ gauss in magnetars.Moreover,cosmic microwave background observations \cite{Planck:2015,Planck:2018,Sutton:2017,Jedamzik:2018} introduce upper and lower bounds.In addition,gamma rays emitted by distant blazars \cite{Neronov:2010,Tavecchio:2010,Taylor:2011,Caprini:2015} have shown the strength of large-scale magnetic fields $ B_{0} $ ranging from  $ 10^{-17} $ to $ 10^{-9} $ gauss. 

In order to understand the origins of these magnetic fields , several investigations have been studied in literature such as the theory of structure formation through the astrophysical Biermann battery mechanism \cite{Biermann:1950}.In this mechanism , these fields are then amplified through various forms of dynamo and then spread into  the intergalactic medium by outflows from galaxies. \cite{Zeldovich:1980book,Lesch:1995,Kulsrud:1997,Colgate:2001,Rees:1987,Daly:1990,Ensslin:1997,Bertone:2006}.

Another theory implies that the origin of these fields is primordial and produced in early Universe \cite{Turner:1988,Ratra:1992,Hogan:1983,Quashnock:1989,Vachaspati:1991}.Among these theories ,it is thought that the most natural mechanism for the generation of large-coherence-scale magnetic fields would be inflation, a period of rapid expansion in the early Universe \cite{Turner:1988}.

In studies of early Universe many authors have  indicated that during inflation ,quantum fluctuation of massless scalar and tensor fields can be amplified significantly which is thought to led to the formation of the large-scale structures observed in Universe today\cite{Mukhanov:1981,Hawking:1982,Starobinsky:1982,Guth:1982,Bardeen:1983}.In addition, it is thought that this amplification is responsible for the generation of relic gravitational waves \cite{Grishchuk:1975,Starobinsky:1979,Rubakov:1982}.

However,the conformal invariance of the Maxwell action does not allow of generation of any large-scaled magnetic fields  \cite{Parker:1968}.Therefore, in this paper we break the conformal invariance by axial coupling interaction term of the form $ I\left(\phi\right)F_{\mu\nu}\tilde{F}^{\mu\nu} $ \cite{Dolgov:1993,Gasperini:1995,Giovannini:2000,Atmjeet:2014} where $ I\left(\phi\right) $ is  a coupling function of the inflaton field  $ \phi $ and $ F_{\mu\nu}=\partial_{\mu} A_{\nu}-\partial_{\nu}A_{\mu} $ is the electromagnetic field tensor and $ \tilde{F}^{\mu\nu}=\frac{1}{2}\frac{\epsilon^{\mu\nu\rho\sigma}}{\sqrt{-g}}F_{\rho\sigma} $ .In this method , $ \tilde{F}^{\mu\nu} $ does not depend on metric and does not contribute to the energy-momentum tensor.Because $ F_{\mu\nu}\tilde{F}^{\mu\nu}=-4\textbf{E}\cdot\textbf{B} $ , the term $ \textbf{E}\cdot\textbf{B} $  appears in our equations and should be approximated by simpler relation in order to solve  system of equations. More importantly, due to axial coupling of inflaton field and electromagnetic field , the term $ \textbf{E}\cdot\textbf{B} $  implies that the produced magnetic field is helical.Additionally, in axial coupling the electric energy density is almost equal to the magnetic energy density , i.e. $ \rho_{E}\sim \rho_{B} $ \cite{Figueroa:2018,Notari:2016,Fujita:2015,Kamarpour:2021,Kamarpour:2022,Kamarpour:2023-I}.Also, the conformal invariance can be broken by $ I^{2}\left(\phi\right)F_{\mu\nu}F^{\mu\nu} $ which is called kinetic coupling, first introduced by Ratra \cite{Ratra:1992} and discussed in references \cite{Giovannini:2001,Bamba:2004,Martin:2008,Demozzi:2009,Kanno:2009,Ferreira:2013,Ferreira:2014,Vilchinskii:2017}.

In this paper we consider the Schwinger effect in natural inflation model by axial coupling only for strong field regime because for weak field the Schwinger effect is negligible\cite{Sobol-Gorbar:2021}.Briefly , the Schwinger effect is producing of charged particles from vacuum by strong electric field \cite{Schwinger:1951} .

Strong  generated electric field is the result of coupling of the electromagnetic field and inflaton field. This effect has been studied  in several papers.For instance, the case of a constant and homogeneous electric field in  de~Sitter  space-time can be found in Refs. \cite{Afshordi:2014,Froeb:2014,Bavarsad:2016,Stahl:2016a, Stahl:2016b,Hayashinaka:2016a,Hayashinaka:2016b,Sharma:2017, Tangarife:2017,  Hayashinaka:2018, Hayashinaka:thesis, Stahl:2018, Geng:2018, Bavarsad:2018}.

It should be noted that,the cosmological Schwinger effect is consideration of expansion of the Universe and investigates the Schwinger effect in de~Sitter space-time.In this case , expansion of the Universe is exponential.This effect drives some expressions for the production of charged particles by strong electric field.The interesting features of the  cosmological Schwinger effect such as infrared hyper-conductivity in the bosonic with very small mass or massless particles have been studied in  Refs.\cite{Afshordi:2014,Hayashinaka:2016a,Hayashinaka:2016b,Hayashinaka:2018,Hayashinaka:thesis,Stahl:2018}.

However, the constant and homogeneous electric filed for the Schwinger effect contradicts the second  law of thermodynamics because it would require the existence of ad hoc currents \cite{Giovannini:2018a} . Instead, expressions for the Schwinger effect must be used in the case of a time-dependent electric field in the strong-field regime \cite{Kitamoto:2018}.

In this paper, we investigate the Schwinger effect by axial coupling in the natural inflation model with two coupling functions.  The paper is organized as follows: we determine  the model and find the solution for the background equations in the natural inflation model in Sect.~\ref{sec-inflation}, where we also consider the axial coupling of the inflation field to the electromagnetic field with two coupling functions. We then obtain the mode-function and estimate the range of parameters for which the back-reaction problem does not occur for our model. In this part , we consider a system of self-consistent equations, including the Schwinger effect. Section~\ref{sec-Schwinger} discusses the main idea of the Schwinger effect, including the Schwinger source term. In Sect.~\ref{sec-Numerical}, we perform numerical calculations for both scenarios, when we use the standard  horizon scale $ k_{H}=aH $  and  with the scale  at which a given Fourier begins to become tachyonically unstable  for two coupling functions and compare them. The summary of the obtained results is given in Sect.~\ref{sec-conclusion}.

\section{Natural inflation}
\label{sec-inflation}
We use the potential of the natural inflation model that was proposed in Refs.~\cite{Freese:1990,Adams:1993}. For more details see our previous works in Refs.~\cite{Kamarpour-Sobol:2018,Kamarpour:2023}.  
\begin{equation}
\label{potential}
V\left(\phi\right)=\Lambda^{4}\left[1-\cos\left(\frac{\phi}{f}\right)\right]
\end{equation}

We consider a spatially flat Friedmann--Lema\^{i}tre--Robertson--Walker(FLRW) Universe with metric tensor
\begin{equation}
\label{metric}
g_{\mu\nu}={\rm diag}\,(1,\,-a^{2},\,-a^{2},\,-a^{2}), \quad \sqrt{-g}=a^{3},
\end{equation}
and use the natural system of units where $\hbar=c=1$, $M_{p}=(8\pi G)^{-1/2}=2.4\cdot 10^{18}\,{\rm GeV}$ is a reduced Planck mass , and $ e=\sqrt{4\pi\alpha}\approx 0.3 $ is the absolute value of the electron's charge.

\subsection{Action}
The action of the inflation field interacting with electromagnetic field by axial coupling reads
\begin{equation}
\label{action-1}
S=\int d^{4}x \sqrt{-g}\left[\frac{1}{2}\partial^{\mu}\phi\partial_{\mu}\phi-V(\phi)-\frac{1}{4}F_{\mu\nu}F^{\mu\nu}+\frac{1}{4}I\left(\phi\right)F_{\mu\nu}\tilde{F}^{\mu\nu}\right]
\end{equation}

Variation for inflaton field $\phi $ reads
\begin{equation}
\label{motion}
\frac{1}{\sqrt{-g}}\partial{_{\mu}}\left[\sqrt{-g}\partial{^{\mu}\phi}\right]+\frac{dV}{d\phi}
=\frac{1}{4}\frac{dI\left(\phi\right)}{d\phi}F_{\mu\nu}\tilde{F}^{\mu\nu}
\end{equation}
Also variation of gauge field $ A_{\mu} $ gives following relation
\begin{equation}
\label{Maxwell-Tilde}
\frac{1}{\sqrt{-g}}\partial_{\mu}\left[\sqrt{-g}I(\phi)\tilde{F}^{\mu\nu}-\sqrt{-g}F^{\mu\nu}\right]=0
\end{equation}
In above equation $ \tilde{F}^{\mu\nu}=\frac{1}{2}\frac{\epsilon^{\mu\nu\rho\sigma}}{\sqrt{-g}}F_{\rho\sigma} $ in which $ \epsilon^{\mu\nu\rho\sigma} $ is the totally antisymmetric Levi-Civita symbol with $ \epsilon^{0123}=1 $.

Further manipulations of equation(\ref{Maxwell-Tilde}) gives following useful equation
\begin{equation}
\label{Maxwell-}
\frac{1}{\sqrt{-g}}\partial_{\mu}\left[\sqrt{-g}{F}^{\mu\nu}\right]-I^{\prime}\left(\phi\right)\partial_{\mu}\phi \tilde{ F}^{\mu\nu}=0
\end{equation}
where $ I^{\prime}\left(\phi\right)=\frac{dI\left(\phi\right)}{d\phi} $.Another useful equation is given by following relation
\begin{equation}
\label{Maxwell-tilde}
\frac{1}{\sqrt{-g}}\partial_{\mu}\left[\sqrt{-g}\tilde{F}^{\mu\nu}\right]=0
\end{equation}
In Eq. (\ref{motion}) $ F_{\mu\nu}\tilde{F}^{\mu\nu}=-4 \textbf{E}\cdot\textbf{B} $ , by using this equation we obtain
\begin{equation}
\label{Back-reaction}
\ddot{\phi}+3H\dot{\phi}+\frac{dV\left(\phi\right)}{d\phi}=-I^{\prime}\left(\phi\right)\textbf{E}\cdot\textbf{B}
\end{equation} 

In obtaining the above equation we assume homogeneous  inflaton field $ \phi=\phi\left(t\right) $.

Now, if we add a following  gauge invariant Lagrangian  $ L_{charged}\left(A,\chi\right)  $ to the action of equation \ref{action-1}
\begin{equation}
S_{gauge}=\int d^{4}x\sqrt{-g}L_{charged}\left(A,\chi\right),
\end{equation}
then the total action will be  given by following relation
\begin{equation}
\label{Total-Action}
S+S_{gauge}=\int d^{4}x \sqrt{-g}\left[\frac{1}{2}\partial^{\mu}\phi\partial_{\mu}\phi-V(\phi)-\frac{1}{4}F_{\mu\nu}F^{\mu\nu}+\frac{1}{4}I\left(\phi\right)F_{\mu\nu}\tilde{F}^{\mu\nu}+L_{charged}\left(A,\chi\right)\right]
\end{equation}
By variation of above action with respect to $ A_{\mu} $ we find
\begin{equation}
\label{Maxwell-tilde-2}
\frac{1}{\sqrt{-g}}\partial_{\mu}\left[\sqrt{-g}{F}^{\mu\nu}\right]-I^{\prime}\left(\phi\right)\partial_{\mu}\phi \tilde{ F}^{\mu\nu}=-j^{\nu}
\end{equation}

In above equation $ j^{\mu}$ is given by
\begin{equation}
J^{\mu}=\frac{\partial L_{charged}\left(A,\chi\right)}{\partial A_{\mu}}
\end{equation}
We use Coulomb gauge for electromagnetic field , i.e. $ A_{\mu}=\left(0,\textbf{A}\right) $ and $ \nabla\cdot\textbf{A}=0    $ .

Electric and magnetic fields are given by following relations
\begin{equation}
\label{Electric-Magnetic}
\textbf{E}=-\frac{1}{a}\dot{\textbf{A}},\hspace{.5cm}\textbf{B}=\frac{1}{a^{2}}\nabla\times\textbf{A}
\end{equation}
In equation \ref{Electric-Magnetic} , $ a=a\left(t\right) $ is scale factor of FLRW Universe.In terms of electromagnetic field tensor $ F^{\mu\nu}=\partial^{\mu}A^{\nu}-\partial^{\nu}A^{\mu} $ and its dual  tensor $ \tilde{F}^{\mu\nu} $ the components of electric and magnetic fields are given by following relations
\begin{equation}
\label{EM-tensor}
F^{0i}=\frac{1}{a}E^{i}\hspace{.5cm},F_{ij}=a^{2}\epsilon_{ijk}B^{k}\hspace{.5cm},\tilde{F}^{0i}=\frac{1}{a}B^{i},\hspace{.5cm}\tilde{F}_{ij}=-a^{2}\epsilon_{ijk}E^{k}
\end{equation}
Note that $\epsilon_{ijk} $ is three dimensional Levi-Civita symbol and $ i,j,k  $ indicate components of 3-vectors. By using the components of electric and magnetic fields , i.e.Eqs. (\ref{EM-tensor}) and Eqs.(\ref{Maxwell-tilde} ,  \ref{Maxwell-tilde-2}) we can write system of closed equations.
\begin{equation}
\label{Electric-Current}
\dot{\textbf{E}}+2H\textbf{E}-\frac{1}{a}\nabla\times\textbf{B}-I^{\prime}\left(\phi\right)\dot{\phi}\textbf{B}=-a\textbf{J},
\end{equation} 
\begin{equation}
\label{Magnetic}
\dot{\textbf{B}}+2H\textbf{B}+\frac{1}{a}\nabla\times\textbf{E}=0,
\end{equation}
\begin{equation}
\nabla\cdot\textbf{B}=0,\hspace{.5cm}\nabla\cdot\textbf{E}=0
\end{equation}
Note that in equation (\ref{Electric-Current}) , current $ \textbf{J} $ of charged particles can be written in terms of the generalized conductivity $ \sigma $ , so we have
\begin{equation}
\label{Current-Counductivty}
\textbf{J}=\frac{1}{a}\sigma\textbf{E}
\end{equation}

In oder to close the equations we need to obtain required relations for electric and magnetic energy densities.It is more convenient to introduce energy-momentum tensor.
\begin{equation}
\label{Energy-Momentum}
T_{\mu\nu}=\frac{2}{\sqrt{-g}}\frac{\delta S}{\delta g^{\mu\nu}}=\partial_{\mu}\phi\partial_{\nu}\phi-g^{\alpha\beta}F_{\mu\alpha}F_{\nu\beta}-g_{\mu\nu}L_{0}+T^{charged}_{\mu\nu}
\end{equation} 
In above equation we introduce  $ L_{0}= \frac{1}{2}\partial^{\mu}{\phi}\partial_{\mu}\phi-V\left(\phi\right)-\frac{1}{4}F_{\mu\nu}F^{\mu\nu} $ . As we know $ F_{\mu\nu}\tilde{F}^{\mu\nu}  $ does not appear in  the energy-momentum relation because it does not depend on metric. Now we find energy density from $ T_{00} $ component.
\begin{equation}
\label{Total-Density}
\rho=\frac{1}{2}\dot{\phi}^{2}+V\left(\phi\right)+\frac{1}{2}\left(E^{2}+B^{2}\right)+\rho_{\chi}=\rho_{inf}+\rho_{EM}+\rho_{\chi}
\end{equation}
In equation (\ref{Total-Density}) $ \rho_{EM} $ is the energy density of electromagnetic filed and $ \rho_{\chi} $ is  the energy density of produced charged particles due to the Schwinger effect.Therefore, the Friedmann equation can be written by
\begin{equation}
\label{Friedmann-2}
H^{2}=\frac{1}{3M_{p}^{2}}\left(\rho_{inf}+\rho_{EM}+\rho_{\chi}\right)
\end{equation} 
Let us look at equation(\ref{Back-reaction}).In this equation the right hand side is back-reaction term and it demonstrates helical nature of electromagnetic field.In addition ,this term should be approximated by simpler relation such as $ \rho_{EM} $ in order to obtain system of closed equation for numerical calculations.We will come back to this equation in section of numerical calculations and discuss about it. 

It is more convenient to write equations (\ref{Electric-Current} , \ref{Magnetic}) in terms of energy density.By using equations (\ref{Electric-Current} , \ref{Current-Counductivty}) and (\ref{Magnetic}) we find
\begin{equation}
\label{EM-Density}
\dot{\rho}_{EM}+4H\rho_{EM}+ 2\sigma\rho_{E}-I^{\prime}\left(\phi\right)\dot{\phi}\textbf{E}\cdot\textbf{B}=0
\end{equation}
In equation (\ref{EM-Density}) the term $ 2\sigma\rho_{E} $ indicates dissipation of the electromagnetic energy density due to the Schwinger effect.We will return to this equation in numerical calculations.As we discussed before about $ \textbf{E}\cdot\textbf{B} $ the term $ I^{\prime}\left(\phi\right)\dot{\phi}\textbf{E}\cdot\textbf{B} $ describes axial coupling nature of electromagnetic field and inflaton field.In fact , this term implies transfer of energy density from inflaton field to electromagnetic field.
\begin{figure}[ht]
	\centering
	\includegraphics[width=0.45\textwidth]{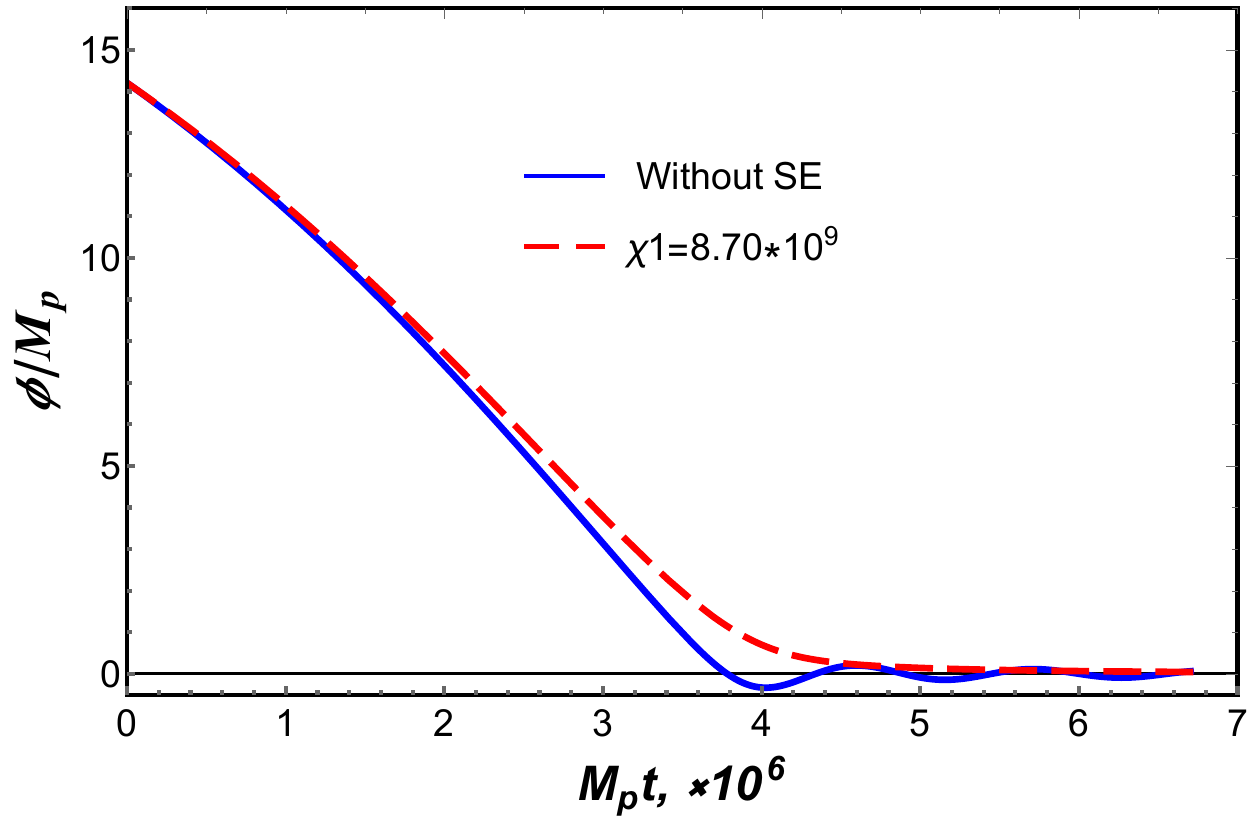}
	\hspace*{2mm}
	\includegraphics[width=0.45\textwidth]{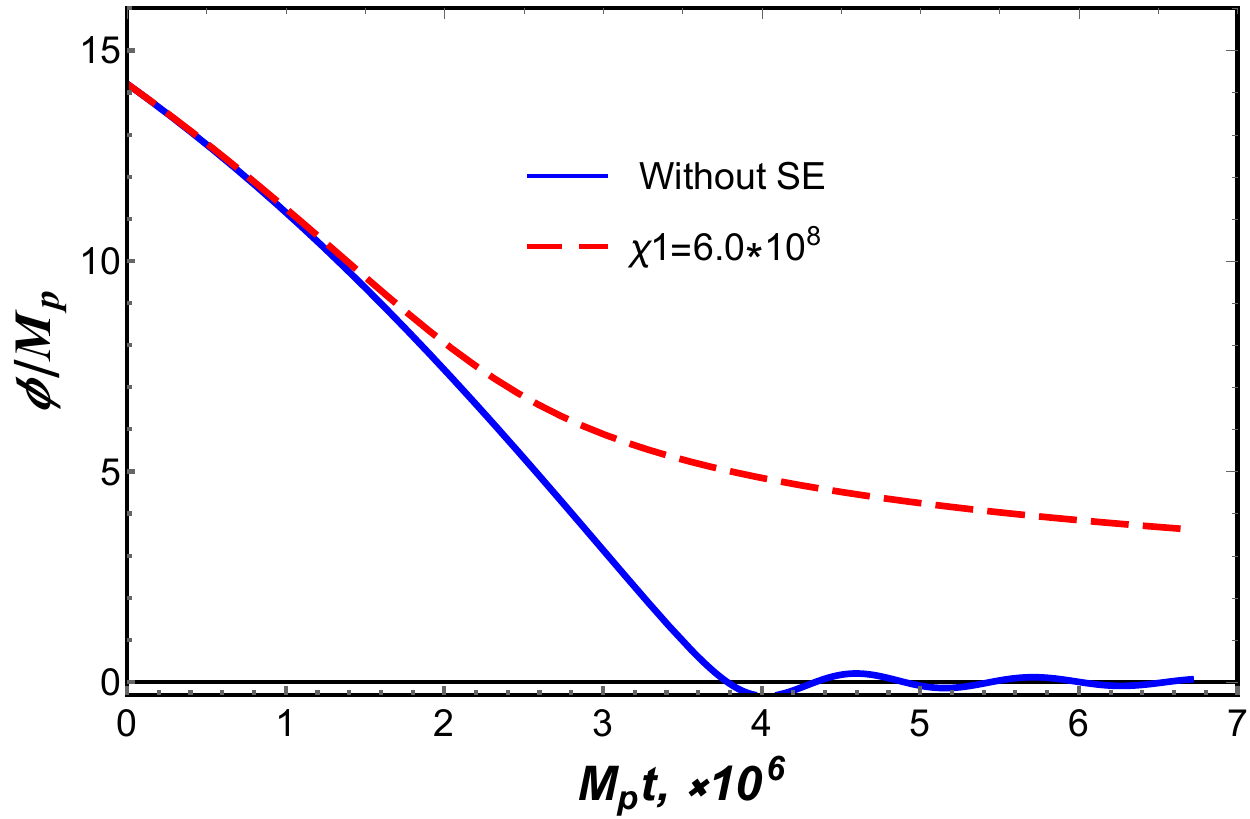}
	\caption{The time dependence of  inflaton field for the simplest coupling function    Eq.\ref{Coupling-1} (a) and  for non-minimal coupling to gravity  Eq.\ref{Coupling-2} (b)  for different values of parameter $ \chi_{1} $.Both without the Schwinger effect and with the Schwinger effect.In both figures it is obvious that strong coupling or back-reaction  occurs.But in panel (b) we see that for non-minimal coupling to gravity back-reaction is much stronger.Note that in obtaining these figures we use Eq.\ref{Last-1} for boundary term, i.e. $ k_{H}=aH $.  }
	\label{Inflaton-1}
\end{figure}
\begin{figure}[ht]
	\centering
	\includegraphics[width=0.45\textwidth]{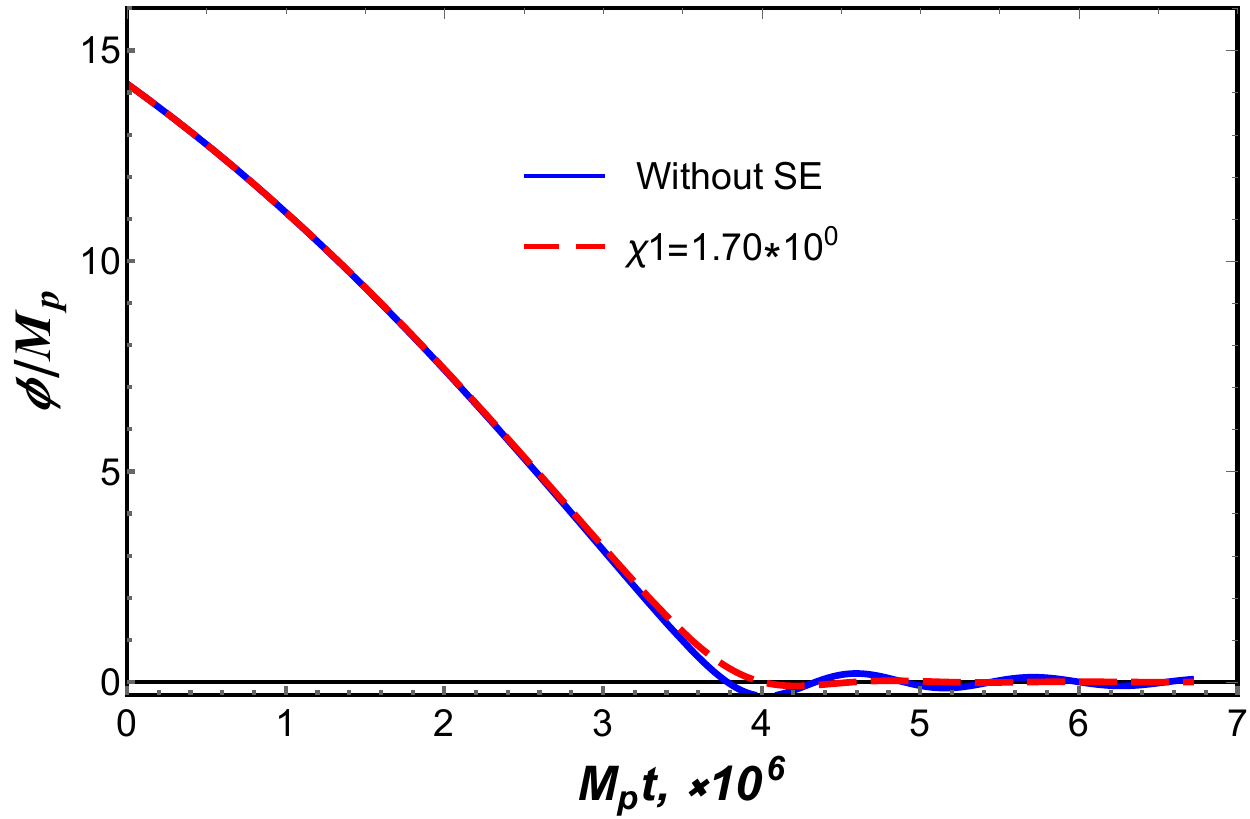}
	\hspace*{2mm}
	\includegraphics[width=0.45\textwidth]{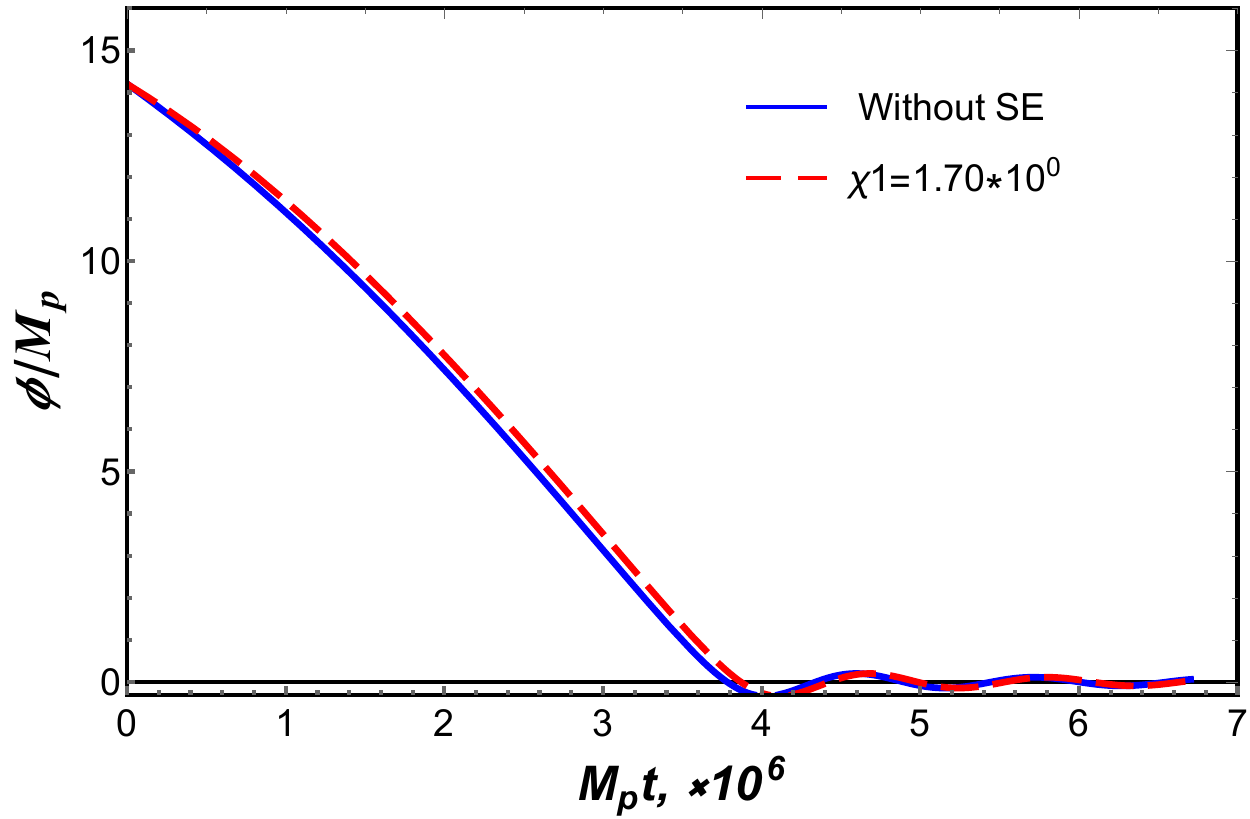}
	\caption{The time dependence of inflaton field for the simplest coupling function  Eq.\ref{Coupling-1} (a) and for non-minimal coupling to gravity  Eq.\ref{Coupling-2}  (b)  for the same values of parameter $ \chi_{1} $.Both without the Schwinger effect and with the Schwinger effect.In panel (a) it seems  that strong coupling or back-reaction does not occur.But in panel (b) we see that for non-minimal coupling to gravity back-reaction occurs and it  is much stronger.Note that in obtaining these figures we use Eq.\ref{Last-2} for boundary term $ K_{H}=aH|\zeta|$ .The effect of Eq.\ref{Last-2} is reducing the value of coupling constant $ \chi_{1}$ and as we see latter in Figs of energy densities  for both panels back-reaction occurs.  }
	\label{Inflaton-2}
\end{figure}
\begin{figure}[ht]
	\centering
	\includegraphics[width=0.45\textwidth]{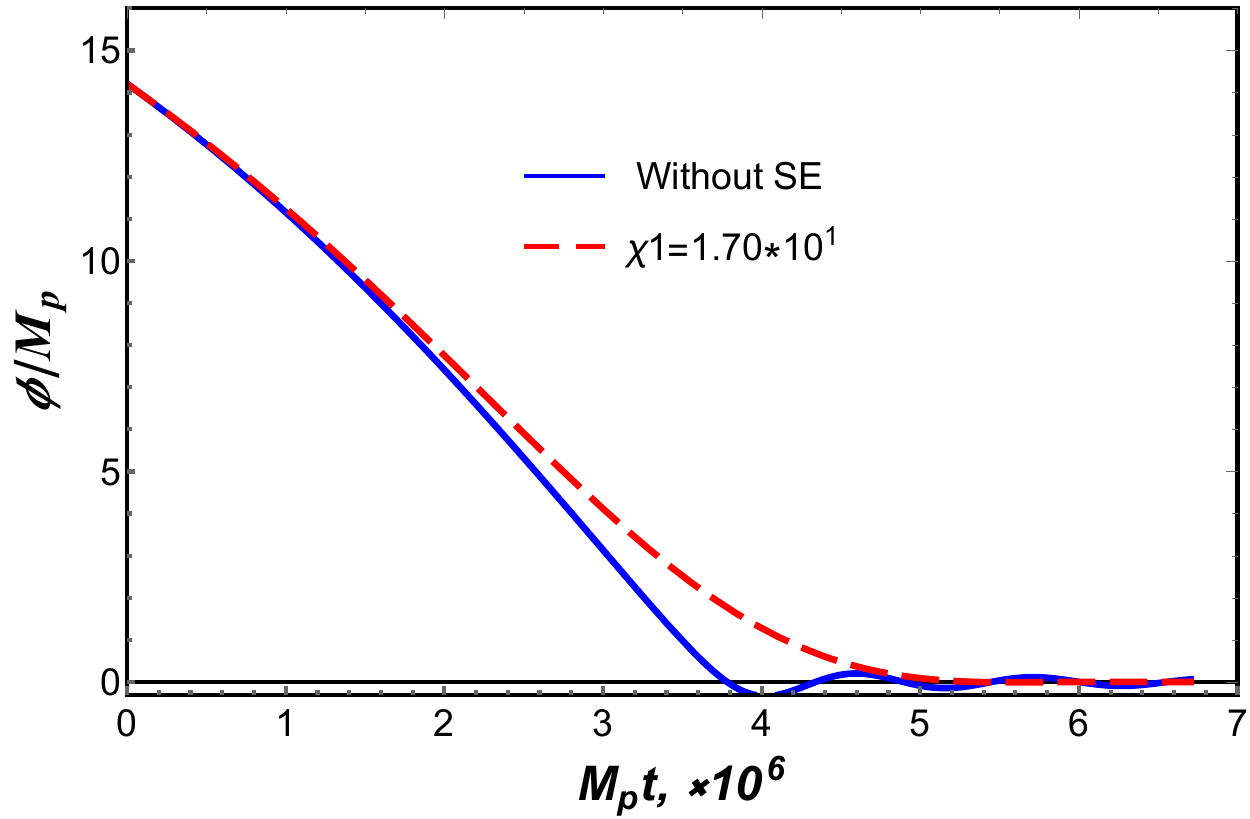}
	\hspace*{2mm}
	\includegraphics[width=0.45\textwidth]{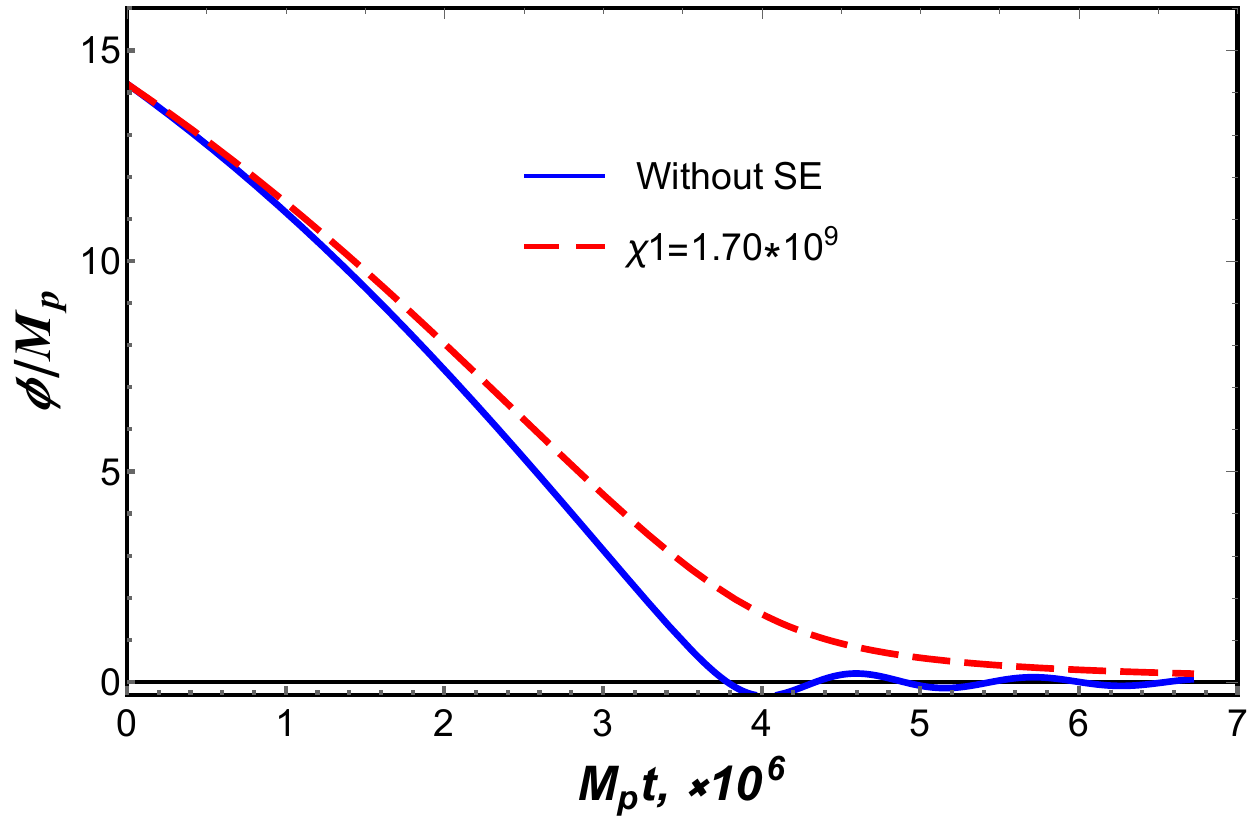}
	\caption{The time dependence of  inflaton field for the simplest coupling function  Eq.\ref{Coupling-1} (a) and  for the same coupling  Eq.\ref{Coupling-1} (b)  for different values  of parameter $ \chi_{1} $.Both without(blue) the Schwinger effect and with (red)the Schwinger effect.In panel (a) it seems  that strong coupling or back-reaction is weaker than panel (b).But in panel (b) we see that $ \chi_{1} $ is much greater than coupling constant in panel a. Note that in obtaining panel (a) we use Eq. \ref{Last-2} $ k_{H}=aH|\zeta| $ for boundary term , whereas in panel (b) we use Eq.\ref{Last-1} $ k_{H}=aH $.The effect of Eq. \ref{Last-2} $ k_{H}=aH|\zeta| $ in panel (a)  is reducing the value of coupling constant $ \chi_{1}$.In fact, when we use tachyonic instability for $ k_{H} $ the effect appears both in values of coupling constant and in strong coupling problem.We see that strong coupling problem in panel (a) is weaker than panel (b).   }
	\label{Inflaton-3}
\end{figure}
\subsection{Mode function}

Let us look at action (\ref{Total-Action}).Terms $ -\int d^{4}x\sqrt{-g}F_{\mu\nu}F^{\mu\nu} $ and $ \int d^{4}x\sqrt{-g}F_{\mu\nu}\tilde{F}^{\mu\nu}  $can be written in coulomb gauge $ A_{\mu}=\left(A_{0},A_{i}\right) $ with $ A_{i}=A^{T}_{i}+\partial_{i}\chi $ , where $ \partial_{i}A^{T}_{i}=0 $ , then the Maxwell and interaction action read
\begin{equation}
\label{Maxwell-Interaction}
S_{Maxwell}+S_{Interaction}=\frac{1}{2}\int d^{4}x\left({A^{T}_{i}}^{\prime}{A^{T}_{i}}^{\prime}+A^{T}_{i}\Delta A^{T}_{i}\right)+\int d^{4}x I\left(\phi\right)\epsilon_{ijk}{A^{T}_{i}}^{\prime}\partial_{j}A^{T}_{k}
\end{equation}
where $ \sqrt{-g}F_{\mu\nu}\tilde{F}^{\mu\nu}=4\epsilon_{ijk}{A^{T}_{i}}^{\prime}\partial_{j}A^{T}_{k} $ and $ \Delta $ is the Laplacian which is calculated respect to the Euclidean metric and $ \prime $ denotes derivative with respect to the conformal time $ \eta =\int\frac{dt}{a} $ .In addition , $ d^{4}x=d\eta d^{3}x $.

By solving equation (\ref{Maxwell-Interaction}) in coulomb gauge , we achieve following relation
\begin{equation}
\label{Vector-Potential}
{A^{T}_{i}}^{\prime\prime}-\nabla^{2}A^{T}_{i}-I^{\prime}\left(\phi\right)\epsilon_{ijk}{A^{T}_{i}}^{\prime}\partial_{j}A^{T}_{k}=0
\end{equation}
The Fourier mode of equation (\ref{Vector-Potential}) is given by following relation
\begin{equation}
\label{Mode-1}
\mathcal {A}^{\prime\prime}_{\eta}+\left(k^{2}+hkI^{\prime}\left(\phi\right)\right)\mathcal{A}_{\eta}=0
\end{equation}
In terms of cosmic time , the above mode-function is given by following equation
\begin{equation}
\label{mode-3}
\ddot{\mathcal{A}_{h}}\left(t,k\right)+H\dot{\mathcal{A}_{h}}\left(t,k\right)+\left(\frac{k^{2}}{a^{2}\left(t\right)}+h\dot{I}\left(\phi\right)\frac{{k}}{a}\right)\mathcal{A}_{h}\left(t,k\right)=0
\end{equation}
In above equation  $ h=\pm $ shows the helicity.Also , the Fourier modes of the transverse part of vector potential can be written as  $  A^{T}_{i}\left(\eta,\textbf{k}\right)=\mathcal{A}_{+}\varepsilon_{+}+\mathcal{A}_{-}\varepsilon_{-} $.For more details about decomposition function in Fourier  space and orthogonality relations, see our previous works in Refs.\cite{Kamarpour:2021,Kamarpour:2022,Kamarpour:2023-I}.

\section{The Schwinger effect}
\label{sec-Schwinger}
We  only consider expressions in strong field regime.As discussed in Refs.\cite{Kamarpour:2022,Kamarpour:2023,Kamarpour:2023-I,Sobol-Gorbar:2021} the Schwinger effect in weak field regime is quite negligible.Thus we consider only two expressions for numerical calculations.
\begin{equation}
\label{sigma}
\sigma_{s}=\frac{g_{s}}{12\pi^{3}}\frac{e^{3}E}{H}\exp\left({-\frac{\pi m^{2}}{|e E|}}\right), \hspace{1cm} s=b,f
\end{equation} 

In Eq.(\ref{sigma}) , $ g_{b}=1 $ and $ g_{f} $ are the number of spin degrees of freedom(d.o.f.).

The equation which describes created charged particles from vacuum is:
\begin{equation}
\label{charged-e}
\dot{\rho}_{\chi}+4H\rho_{\chi}=2\rho_{E}\sigma_{s}
\end{equation}
In Eq.(\ref{charged-e}) $ 4H $ appears because we neglect mass and only consider massless charged particles.In fact, produced charged particles due to the Schwinger effect have masses smaller than the Hubble parameter.
\begin{figure}[ht]
	\centering
	\includegraphics[width=0.45\textwidth]{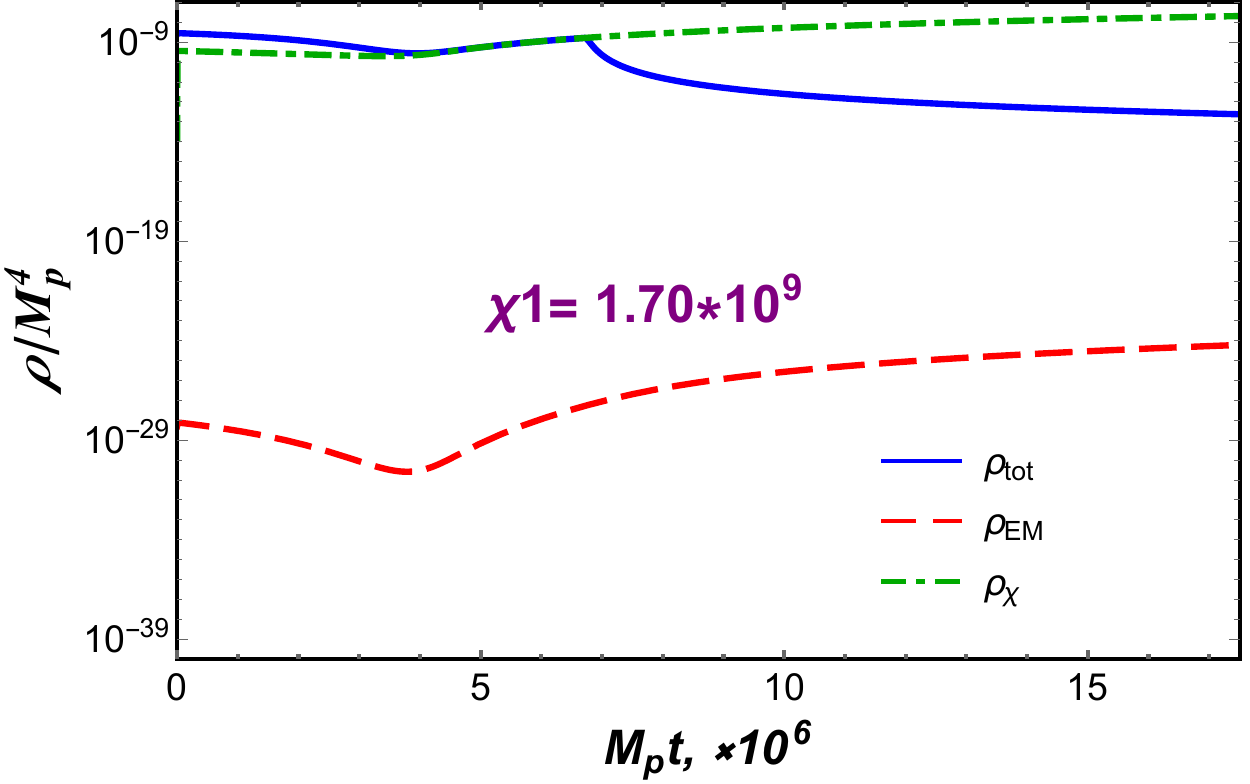}
	\hspace*{2mm}
	\includegraphics[width=0.45\textwidth]{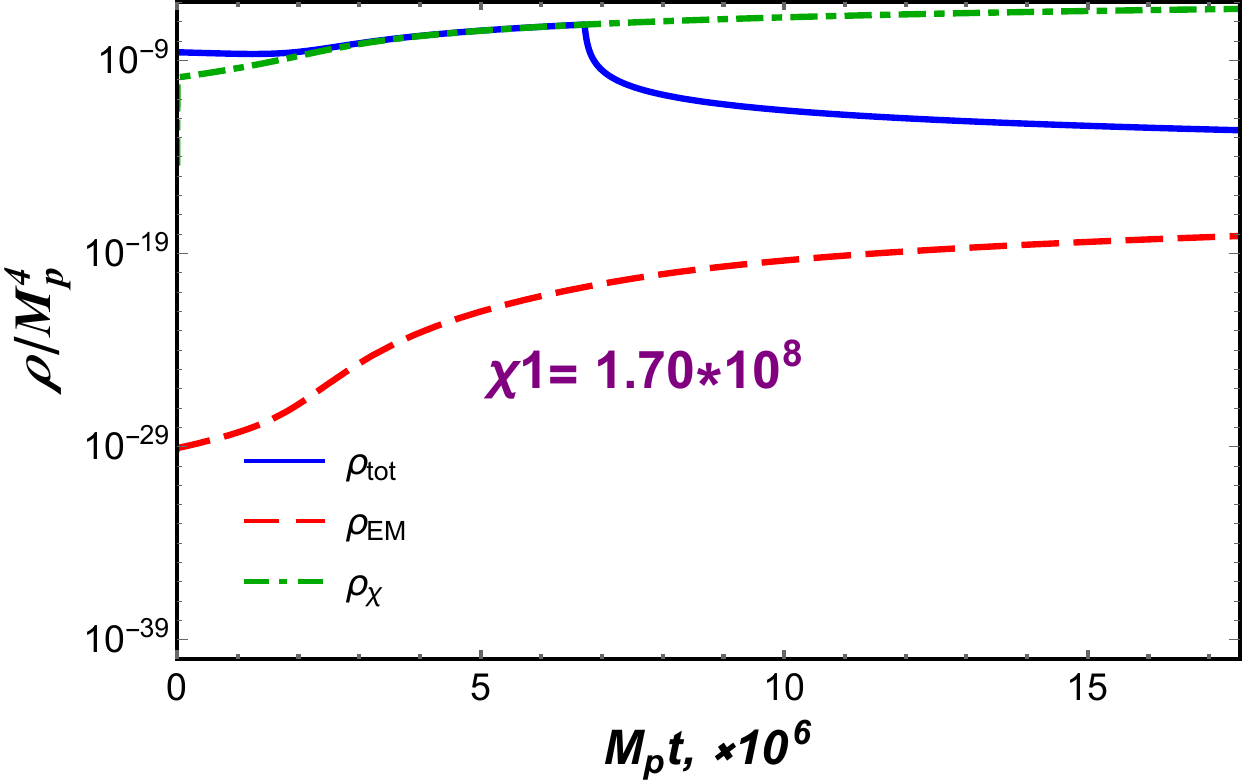}
	\caption{The time dependence (a)  of energy densities  for $ \chi_{1}=1.70\times 10^{9} $ and  the simplest coupling function  Eq.\ref{Coupling-1} and  (b) for  non-minimal coupling to gravity  Eq.\ref{Coupling-2} and $ \chi_{1}=1.70\times 10^{8} $ .In both panels we see strong coupling or back-reaction problem due to produced charged particles. In each panel $ \rho_{tot} $ (blue), $ \rho_{EM}$ (red dashed line), $ \rho_{\chi} $ (green dashed line )  show the total energy density, electromagnetic energy density and energy density of charged particles due to  the Schwinger effect respectively.Note that electromagnetic energy density is small but the density of charged particles is very high which causes back-reaction problem.We use boundary term of Eq.\ref{Last-1} i.e. $ k_{H}=aH $ .    }
	\label{Density-1}
\end{figure}
\begin{figure}[ht]
	\centering
	\includegraphics[width=0.45\textwidth]{SE-Sim-9}
	\hspace*{2mm}
	\includegraphics[width=0.45\textwidth]{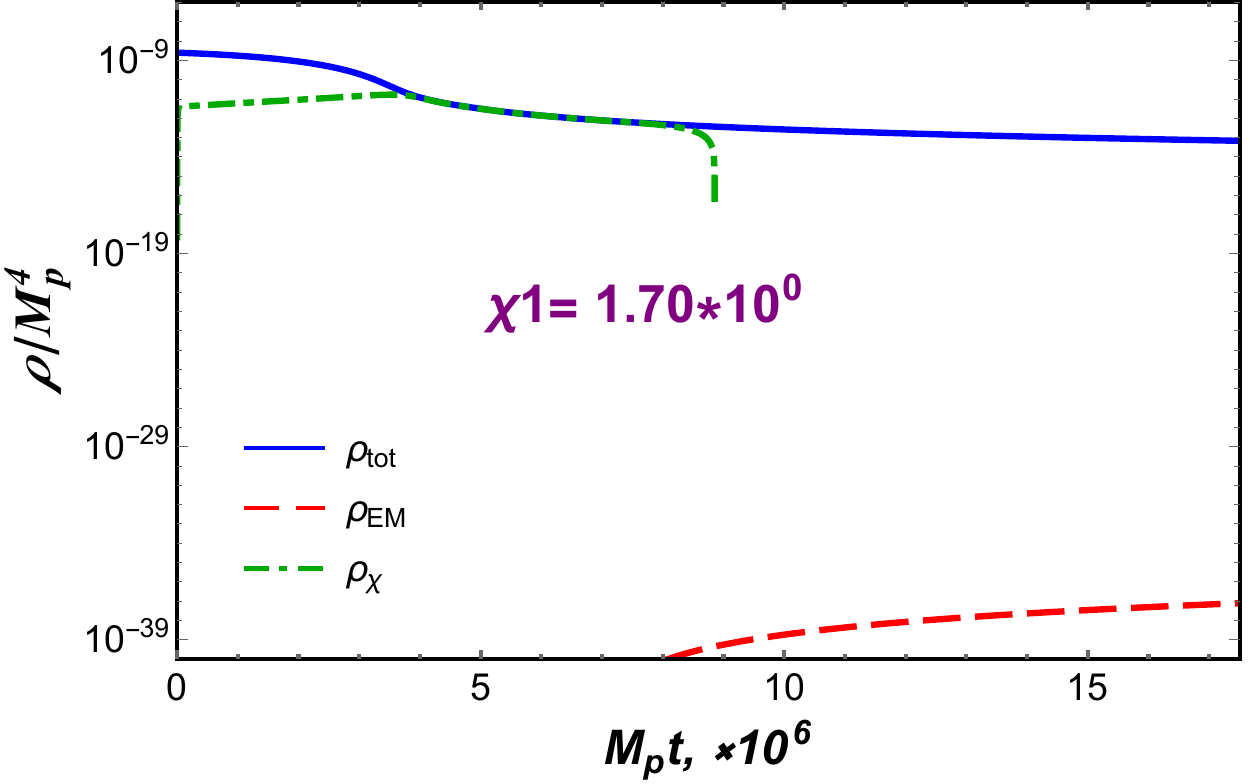}
	\caption{ The time dependence (a)  of energy densities  for $ \chi_{1}=1.70\times 10^{9} $ and the simplest coupling function  Eq.\ref{Coupling-1} and  (b) for the same simplest coupling  function  Eq.\ref{Coupling-1} and $ \chi_{1}=1.70\times 10^{0} $ .In both panels we see strong coupling or back-reaction problem due to produced charged particles. In each panel $ \rho_{tot} $ (blue) , $ \rho_{EM} $ (red dashed line) , $ \rho_{\chi} $ (green dashed line )  show the total energy density, electromagnetic energy density and energy density of charged particles due to  the Schwinger effect respectively.Note that electromagnetic energy density is small but the  energy density of charged particles is very high which causes back-reaction problem.We use boundary term of Eq.\ref{Last-1} $ k_{H}=aH $ for panel (a) and Eq.\ref{Last-2} $ k_{H}=aH|\zeta| $ for panel (b).As we see in Figs(\ref{Inflaton-2}-a -b) and (\ref{Inflaton-3}-a) , when we use tachyonic instability for $ k_{H} $ the value of coupling constant reduces and effect of back-reaction is weaker than the case of standard horizon scale $ k=aH $ .  }
	\label{Density-2}
\end{figure}
\begin{figure}[ht]
	\centering
	\includegraphics[width=0.45\textwidth]{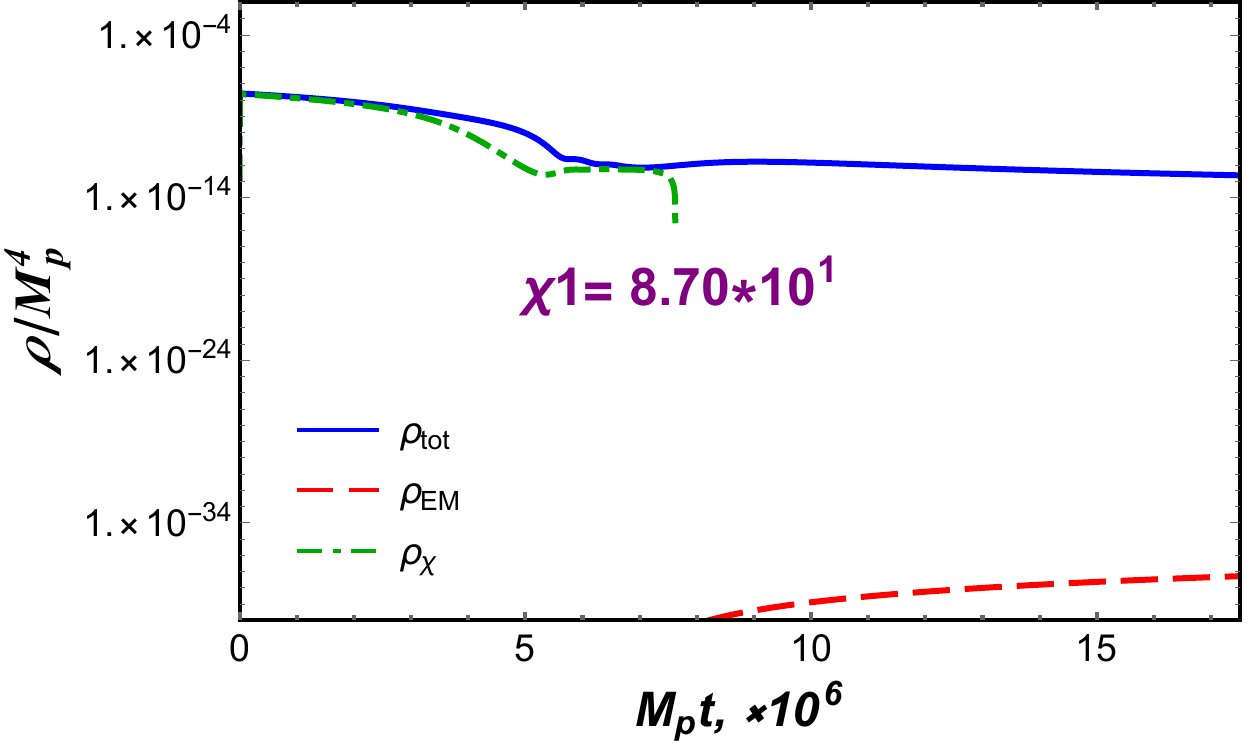}
	\hspace*{2mm}
	\includegraphics[width=0.45\textwidth]{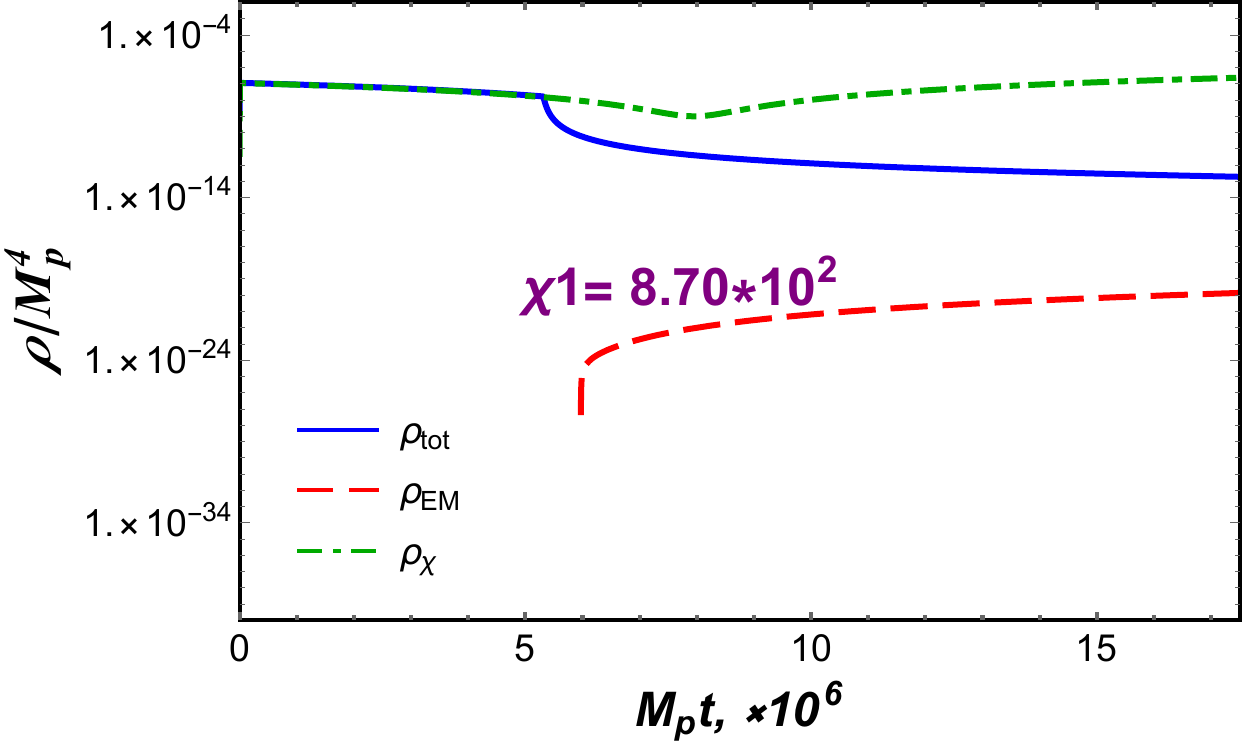}
	\caption{The time dependence (a)  of energy densities  for $ \chi_{1}=8.70\times 10^{1} $ and non-minimal coupling function to gravity  Eq.\ref{Coupling-2} and  (b) for the same  non-minimal coupling to gravity  Eq.\ref{Coupling-2} and $ \chi_{1}=8.70\times 10^{2} $ .In both panels we see strong coupling or back-reaction problem due to produced charged particles. In each panel $ \rho_{tot} $ (blue) , $ \rho_{EM} $ (red dashed line), $\rho_{\chi} $ (green dashed line )  show the total energy density, electromagnetic energy density and energy density of charged particles due to  the Schwinger effect respectively.Note that electromagnetic energy density is small but the density of charged particles is very high which causes back-reaction problem.We use boundary term of Eq.\ref{Last-2} $ k_{H}=aH|\zeta| $.}
	\label{Density-3}
\end{figure}

\begin{figure}[ht]
	\centering
	\includegraphics[width=0.45\textwidth]{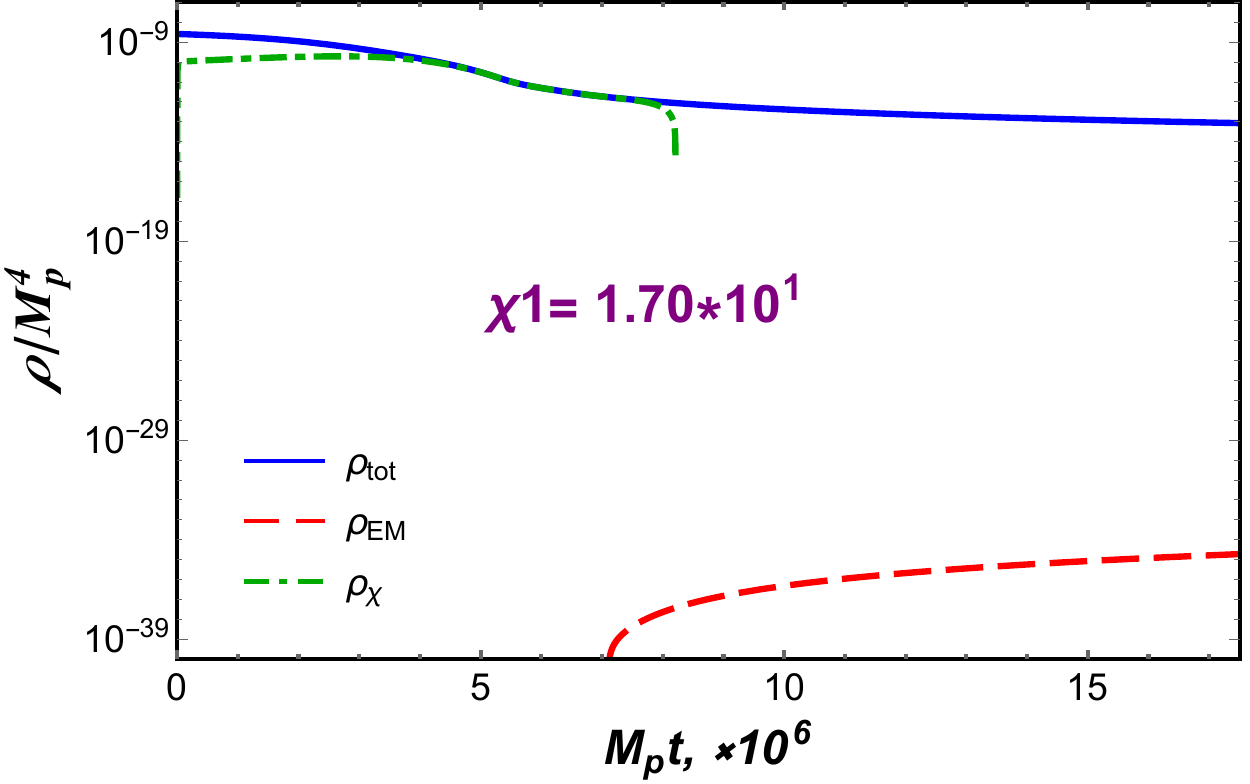}
	\hspace*{2mm}
	\includegraphics[width=0.45\textwidth]{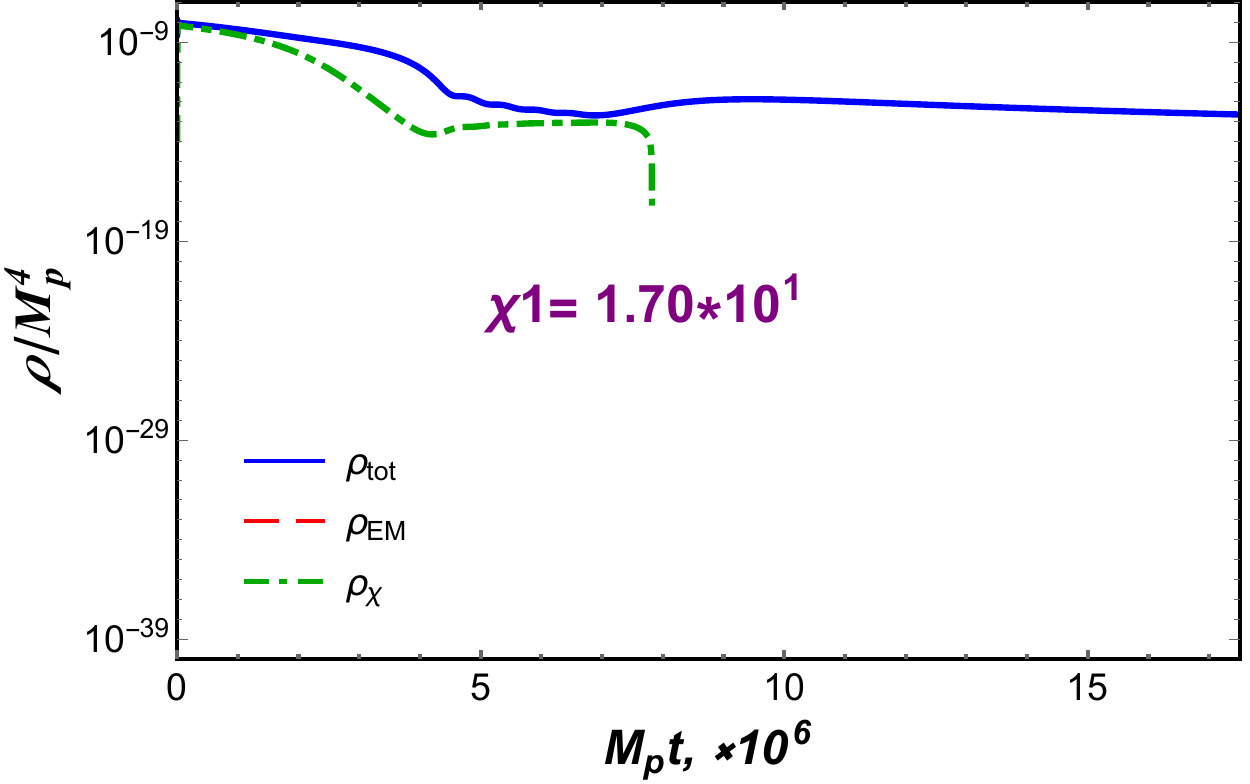}
	\caption{The time dependence (a)  of energy densities  for $ \chi_{1}=1.70\times 10^{1} $ and the  simplest coupling function Eq.\ref{Coupling-1} and  (b) for   non-minimal coupling to gravity Eq.\ref{Coupling-2} and $ \chi_{1}=1.70\times 10^{1} $ .In both panels we see strong coupling or back-reaction problem due to produced charged particles. In each panel $ \rho_{tot} $ (blue) , $ \rho_{EM} $ (red dashed line) , $ \rho_{\chi} $ (green dashed line )  show the total energy density, electromagnetic energy density and energy density of charged particles due to  the Schwinger effect respectively.Note that electromagnetic energy density is small but the density of charged particles is very high which causes back-reaction problem.We use boundary term of Eq.\ref{Last-2} $ k_{H}=aH|\zeta| $.}
	\label{Density-4}
\end{figure}
\section{Numerical calculations}
\label{sec-Numerical}
\subsection{The simplest Coupling function}

Before we start numerical calculations , it should be emphasized that required information about CMB constraints and slow-roll parameters such as spectral index,tensor to scalar ratio and other relevant informations $ \epsilon , \eta,n_{s} , r, $ are given in our previous work of Ref\cite{Kamarpour-Sobol:2018}.  For numerical calculations it is convenient to use the  following simplest coupling function.
\begin{equation}
\label{Coupling-1}
I\left(\phi\right)=\chi_{1}\frac{\phi}{M_{p}}
\end{equation} 
In above relation $ \chi_{1} $ is dimensionless coupling constant.Using this coupling function gives insight for numerical calculations.Let us look at Eq.(\ref{Back-reaction}).If we think of slow-roll condition then one may eliminate $ \ddot{\phi} $ and $ \dot{\phi} $ and obtain following relation
\begin{equation}
\frac{dV\left(\phi\right)}{d\phi}\sim -{I}^{\prime}\left(\phi\right)\textbf{E}\cdot\textbf{B}\sim-\frac{\chi_{1}}{M_{p}}\textbf{E}\cdot\textbf{B}
\end{equation}
 Using slow-roll parameter  $\epsilon=\frac{M_{p}^{2}}{2}\left(\frac{V^{\prime}}{V}\right)^{2}  $ and assuming in slow roll condition $ \rho_{inf}\sim V\left(\phi\right) $ then we find 
 \begin{equation}
 \label{epsilon}
\epsilon\sim\frac{1}{2}\left(\frac{\chi_{1}\textbf{E}\cdot\textbf{B}}{\rho_{inf}}\right)^{2}
 \end{equation}
 When we use approximation for $ \textbf{E}\cdot\textbf{B} $ then the above relation will be useful.See \cite{Sobol-Gorbar:2021}.
 
 \subsection{Non-minimal coupling to gravity}
 
One may choose coupling function $ I\left(\phi\right) $ from conformal transformation $ \tilde{g}_{\mu\nu}=\Omega^{2}g_{\mu\nu} $ and achieve  following relation.See \cite{Kamarpour:2021,Kamarpour:G,Kamarpour:2022,Sobol:2021A}
\begin{equation}
\label{Coupling-2}
I\left(\phi\right)=12\chi_{1}e^{\left(\sqrt{\frac{2}{3}}\frac{\phi}{M_{p}}\right)}\left[\frac{1}{3M_{p}^{2}}\left(4V\left(\phi\right)\right)+\frac{\sqrt{2}}{\sqrt{3}M_{p}}\left(\frac{dV}{d\phi}\right)\right]
\end{equation}
In above equation the coupling constant $\chi_{1}$ has dimension $ M_{p}^{-2} $ .Inserting potential (\ref{potential})  into Eq.(\ref{Coupling-2}) the non-minimal  coupling function is given by following equation
\begin{equation}
\label{Coupling-3}
I\left(\phi\right)=12\chi_{1}e^{\sqrt{\frac{2}{3}}\frac{\phi}{M_{p}}}\left[\frac{4\Lambda^{4}}{3M_{p}^{2}}\left(1-\cos\left(\frac{\phi}{f}\right)\right)+\frac{\sqrt{2}}{\sqrt{3}M_{p}}\frac{\Lambda^{4}}{f}\sin\left(\frac{\phi}{f}\right)\right]
\end{equation}
By taking derivative of Eq.(\ref{Coupling-3}) we find
\begin{equation}
\label{D-Coupling}
\dot{I}\left(\phi\right)=12\dot{\phi}\chi_{1}e^{\sqrt{\frac{2}{3}}\frac{\phi}{M_{p}}}\left[\frac{\sqrt{2}}{\sqrt{3}}\frac{4\Lambda^{4}}{3M_{p}^{2}}\left(1-\cos\left(\frac{\phi}{f}\right)\right)+\frac{2}{3M_{p}^{2}}\frac{\Lambda^{4}}{f}\sin\left(\frac{\phi}{f}\right)+\frac{\sqrt{2}}{\sqrt{3}M_{p}}\frac{\Lambda^{4}}{f^{2}}\cos\left(\frac{\phi}{f}\right)\right]
\end{equation}
In oder to switch on the Schwinger effect we must numerically solve equations (\ref{Back-reaction} , \ref{Friedmann-2} ,\ref{EM-Density} , \ref{charged-e} ) by using Eq.(\ref{sigma}) into Eq.(\ref{charged-e}) and setting $ \textbf{E}\cdot\textbf{B}\sim \sqrt{2\rho_{E}}\sqrt{2\rho_{B}}=2\sqrt{\rho_{E}\rho_{B}}\sim\rho_{EM} $.For this approximation we argue that in axial coupling $ \rho_{E}\sim \rho_{B} $ .See Refs.\cite{Figueroa:2018,Notari:2016,Fujita:2015,Kamarpour:2021,Kamarpour:2022,Kamarpour:2023-I}
\begin{figure}[ht]
	\centering
	\includegraphics[width=0.45\textwidth]{SE-Sim-b-0-1}
	\hspace*{2mm}
	\includegraphics[width=0.45\textwidth]{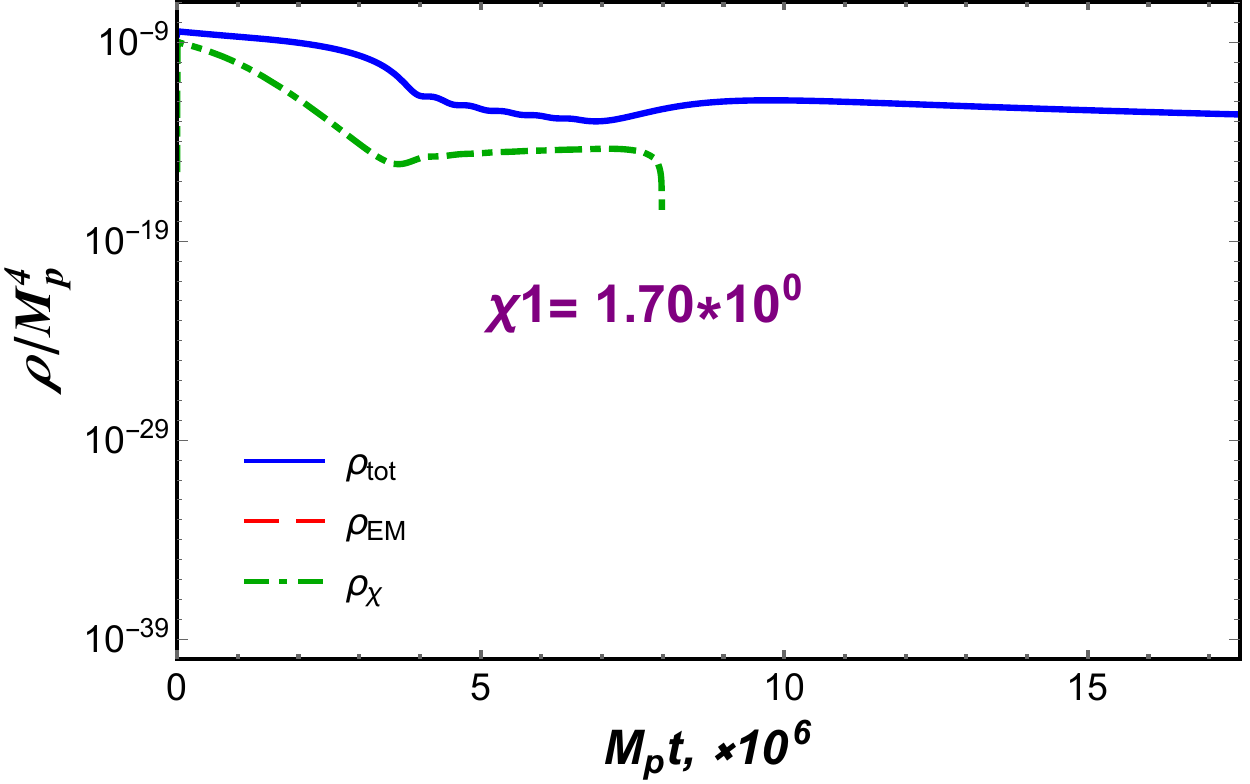}
	\caption{The time dependence (a)  of energy densities  for $ \chi_{1}=1.70\times 10^{0} $ and the  simplest coupling function Eq.\ref{Coupling-1} and  (b) for   non-minimal coupling to gravity Eq.\ref{Coupling-2} and $ \chi_{1}=1.70\times 10^{0} $ .In  panel (a) we see strong coupling or back-reaction problem due to produced charged particles whereas in panel (b) it seems there is no back-reaction problem.Note that in panel (b) there is no electromagnetic field at all.Therefore, for both panels the Schwinger effect  is quite impossible because it spoils inflation and terminates the enhancement of electromagnetic field. In each panel $ \rho_{tot} $ (blue) , $ \rho_{EM} $ (red dashed line) ,  $ \rho_{\chi} $ (green dashed line )  show the total energy density, electromagnetic energy density and energy density of charged particles due to  the Schwinger effect respectively.Note that electromagnetic energy density is small but the density of charged particles is very high which causes back-reaction problem.We use boundary term of Eq.\ref{Last-2} $ k_{H}=aH|\zeta| $}.
	\label{Density-5}
\end{figure}
\subsection{Tachyonic instability}
We should add boundary term to the right hand side of equation (\ref{EM-Density}) but we need to discuss about tachyonic instability.Le us look at Eq.(\ref{mode-3}).

Term in bracket determines tachyonic instability.Tachyonic instability begins when $ h=- $ and $ \dot{I}\left(\phi\right)\frac{k}{a}\geq \frac{k^{2}}{a^{2}} $ or $ \frac{k}{aH}\leq\frac{{I}^{\prime}\left(\phi\right)\dot{\phi}}{H} $.We introduce the parameter $ \zeta $ so that
\begin{equation}
\label{zeta}
  \zeta=\frac{{I}^{\prime}\left(\phi\right)\dot{\phi}}{H} 
\end{equation}
Therefore, the condition for tachyonic instability is $ \frac{k}{aH}\leq|\zeta| $.The critical value for momentum is $ k_{c}=aH|\zeta| $.Thus allowed modes must satisfy $ k<k_{c} $ in order to be detectable.

The above discussion will help us to add required boundary term to the right hand side of Eq.(\ref{EM-Density}).  
\subsection{Initial condition and boundary term}
We use the Bunch-Davies vacuum initial condition for equations (\ref{Mode-1}) and (\ref{mode-3}).
\begin{equation}
\mathcal{A}_{h}\left(t,k\right)=\frac{1}{\sqrt{2k}}e^{-ik\eta}, \hspace{1cm} k\eta\longrightarrow\infty
\end{equation}
Power spectrum of electric field is defined by\cite{Subramanian:2016,Durrer:2013}
\begin{equation}
\label{power-spectrum}
\frac{d\rho_{E}}{d\ln k}=\frac{k^{3}}{\left(2\pi\right)^{2}}\frac{1}{a^{2}}\left(|\frac{\partial\mathcal A_{+}(t,k)}{\partial t}|^{2}+|\frac{\partial\mathcal A_{-}(t,k)}{\partial t}|^{2}\right).
\end{equation} 
Using Eq.(\ref{power-spectrum}) we find required equation for boundary term\cite{Sobol:2018,Kamarpour:2022,Kamarpour:2023,Kamarpour:2023-I}.
\begin{equation}
\label{Last-1}
\left(\dot{\rho}_{E}\right)_{H}=\frac{d\rho_{E}}{dk}\arrowvert_{k=k_{H}}\cdot\frac{dk_{H}}{dt}=\frac{H^{5}}{8\pi^{2}},\hspace{.5cm}k_{H}=aH
\end{equation} 
 One writes Eq.(\ref{Last-1})by setting $ k_{H}=k_{c}=aH|\zeta| $ .Thus the required relation for boundary term is given by\cite{Sobol-Gorbar:2021}.
\begin{equation}
\label{Last-2}
\left(\dot{\rho}_{E}\right)_{H}=\frac{d\rho_{E}}{dk}\arrowvert_{k=k_{H}}\cdot\frac{dk_{H}}{dt}=\frac{H^{5}|\zeta|^{3}}{\pi^{2}},\hspace{.5cm} k_{H}=k_{c}=aH|\zeta|
\end{equation} 
We must emphasize that the Eqs.(\ref{Last-1}) or (\ref{Last-2}) should be added to the right hand side of Eq.(\ref{EM-Density}) separately in order to investigate which one is more appropriate choice.
 
All remains to be done is to solve Eqs.(\ref{Friedmann-2} , \ref{Back-reaction} , \ref{EM-Density} , \ref{charged-e}) and to obtain required results.Before that let us look at figures.

Figure (\ref{Inflaton-1} , \ref{Inflaton-2} , \ref{Inflaton-3}) show the time dependence  of inflation field without Schwinger effect (blue line )and  with Schwinger effect (red dashed line)  for various values of  parameter $ \chi_{1} $. Note that for $ \chi_{1}=1.70\times 10^{0} $ it seems the back-reaction is weak and there is possibility for the Schwinger effect whereas in figure \ref{Density-5} we see the energy density of produced charged particle is high and the Schwinger effect does not occur.In addition in figure \ref{Density-5} electromagnetic field is very small  .

Figure (\ref{Density-1}) shows the time dependence (a)  of energy densities  for $ \chi_{1}=1.70\times10^{9} $ and  (b) for $ \chi_{1}=1.70\times10^{8} $.There is  back-reaction in each panels. In each panel $ \rho_{tot} $ (blue), $ \rho_{_E} $ (red dashed line), $ \rho_{\chi} $ (green dashed line )  show the total energy density, electromagnetic energy density and energy density of charged particles due to Schwinger effect respectively.We identify the horizon scale by $ k_{H}=aH $.

Figure (\ref{Density-2}) shows the time dependence (a)  of energy densities  for $ \chi_{1}=1.70\times 10^{9} $ and  (b) for $ \chi_{1}=1.70\times 10^{0} $. In both panels we see back-reaction due to produced charged particles respectively. In each panel $ \rho_{tot} $ (blue)  ,$ \rho_{_E} $ (red dashed line), $ \rho_{\chi} $ (green dashed line )  show the total energy density, electromagnetic energy density and energy density of charged particles due to Schwinger effect respectively. We see that electromagnetic energy densities are very small  and also energy densities of created charged particles  for these value of parameter $ \chi_{1} $ are high and causes back-reaction .

Figure (\ref{Density-3}) shows the time dependence(a) of energy densities for $ \chi_{1}=1.70\times 10^{1} $ and (b) $ \chi_{1}=8.70\times 10^{2} $.In both panel we use Eq.(\ref{Coupling-2}) for coupling function.In each panel $ \rho_{tot}$ (blue) , $ \rho_{_E} $ (red dashed line), $ \rho_{\chi} $ (green dashed line )  show the total energy density, electromagnetic energy density and energy density of charged particles due to Schwinger effect respectively.Also in both panel we identify the horizon scale by $ k_{H}=aH|\zeta| $.Electromagnetic energy densities in both panels are very small but energy densities of produced charged particles are exceeding of energy density of inflaton field. 

 Figure ( \ref{Density-5} ) shows when Schwinger effect is on there is no electromagnetic filed in panel (b) and in panel(a)  is not significant.
 
 More importantly,let us look at figures (\ref{Inflaton-2}- a) and (\ref{Density-5}- b).It seems back-reaction is weak and the Schwinger effect exists.In both figures $ k_{H}=aH|\zeta| $ in order to avoid back-reaction problem.But we discovered that there is no significant electromagnetic field.Thus the Schwinger effect plays no roles because of strong coupling problem and instead spoils inflaton field.

\section{Conclusion}
\label{sec-conclusion}
In this work we examined the influence of the Schwinger effect on natural inflation model by axial coupling.Our study was divided into two parts.

In first part , we assumed $ \rho_{E}\sim \rho_{B} $ and for this reason we only included $ \rho_{EM} $ in the Friedmann equation (\ref{Friedmann-2}).This assumption was validated by considering axial coupling between electromagnetic field and inflaton field.In addition, the correctness of this  assumption was mentioned and confirmed in literature before \cite{Figueroa:2018,Notari:2016,Fujita:2015,Kamarpour:2021,Kamarpour:2022,Kamarpour:2023-I} .

We used two coupling functions , the simplest coupling of Eq.(\ref{Coupling-1}) and non-minimal coupling to gravity Eq.(\ref{Coupling-2}).In first part ,we incorporated the Schwinger effect in our action and considered the equation of motion for the inflaton field , taking into account the back-reaction term $ \textbf{E}\cdot\textbf{B} $ in the right hand side of equation (\ref{Back-reaction}) and added a gauge invariant action to the action \ref{action-1}.Then we finalized system of closed equations and performed numerical calculations by utilizing the coupling functions (\ref{Coupling-1} , \ref{Coupling-2})  and boundary term (\ref{Last-1}) in relations (\ref{Friedmann-2} , \ref{Back-reaction} , \ref{EM-Density} , \ref{charged-e}).

We produced figures (\ref{Inflaton-1}  , \ref{Inflaton-2} ,\ref{Inflaton-3} , \ref{Density-1} , \ref{Density-2} , \ref{Density-3}  , \ref{Density-4} , \ref{Density-5}) and  observed that when the horizon scale is $ k_{H}=aH $ , back-reaction occurs due to created charged particles.In fact,we found that instantly, the Schwinger effect produces very high energy density of charged particles which causes back-reaction problem.

In second part,we activated another boundary term in order to avoid back-reaction problem.Thus,  we adhered to our assumption of $ \rho_{B} \sim \rho_{E} $, and accordingly, only electromagnetic energy density $ \rho_{EM} $ was included in the Friedmann equation (\ref{Friedmann-2}) and identified  the new horizon scale $ K_{H}=aH|\zeta| , \zeta=\frac{{I}^{\prime}\left(\phi\right)\dot{\phi}}{H} $ .This is the scale at which a given Fourier mode Eq.(\ref{mode-3}) begins to become tachyonically unstable.But choosing this scale  does not alter conclusions of the first part.

Subsequently, we performed numerical calculations  and noticed that ,the effect of choosing tachyonic instability is reducing the values of coupling constant $ \chi_{1} $ and also weakening the back-reaction problem.Therefore, in this investigation we found that the Schwinger effect does not occur because of strong coupling or back-reaction.

Finally,in contrast to our previous works in Refs.\cite{Kamarpour-Sobol:2018,Kamarpour:2023} on natural inflation model in which both magneto-genesis and the Schwinger effect were considerable,in this work we discovered that in axial coupling at least the Schwinger effect is impossible and spoils inflaton field.

One may estimate due to existence of strong back-reaction problem , magneto-genesis by axial coupling in this model is impossible.But this needs separate investigation and should be addressed elsewhere.

\begin{acknowledgments}
	
	The author would like to express  gratitude to S. Vilchinskii, E.V. Gorbar, and O. Sobol for their valuable insights and discussions during the preparation of this manuscript. Special thanks are also extended to O. Sobol for his assistance in creating the figures presented in this paper.

\end{acknowledgments}
 
\section*{Data Availability Statement}	

The Author confirms that this manuscript has no associated data in a data repository.

\section*{Declaration of competing interest }
The authors declare that they have no known competing financial interests or personal relationships
that could have appeared to influence the work reported in this paper.


\end{document}